\documentclass[showpacs,preprintnumbers,amsmath,twocolumn,amssymb]{revtex4}
\usepackage{graphicx}
\usepackage{dcolumn}
\usepackage{bm}
\usepackage{amssymb}
\usepackage{array}
\usepackage{hhline}
\usepackage{longtable}

\begin{document}

\title{Exact solution of mean geodesic distance for Vicsek fractals}

\author{Zhongzhi Zhang$^{1,2}$}
\email{zhangzz@fudan.edu.cn}

\author{Shuigeng Zhou$^{1,2}$}
\email{sgzhou@fudan.edu.cn}

\author{Lichao Chen $^{1,2}$}

\author{Ming Yin$^{1,2}$}

\author{Jihong Guan$^{3}$}

\affiliation {$^{1}$Department of Computer Science and Engineering,
Fudan University, Shanghai 200433, China}

\affiliation {$^{2}$Shanghai Key Lab of Intelligent Information
Processing, Fudan University, Shanghai
200433, China} %

\affiliation{$^{3}$Department of Computer Science and Technology,
Tongji University, 4800 Cao'an Road, Shanghai 201804, China}


\begin{abstract}
The Vicsek fractals are one of the most interesting classes of
fractals and the study of their structural properties is important.
In this paper, the exact formula for the mean geodesic distance of
Vicsek fractals is found. The quantity is computed precisely through
the recurrence relations derived from the self-similar structure of
the fractals considered. The obtained exact solution exhibits that
the mean geodesic distance approximately increases as an exponential
function of the number of nodes, with the exponent equal to the
reciprocal of the fractal dimension.  The closed-form solution is
confirmed by extensive numerical calculations.
\end{abstract}

\date{\today}
\maketitle

The concept of fractals plays an important role in characterizing
the features of complex systems in nature, since many objects in the
real world can be modeled by fractals~\cite{Ma82}. In the last two
decades, a great deal of activity has been concentrated on the
studies of fractals~\cite{HaBe87,BeHa00}. It has been
shown~\cite{BeOs79,GeMaAh80,GeAhMaKi81,GrKa82,ScSc89,HiBe06,Hi07,RoAv07}
that regular fractals capture important aspects of critical
percolation clusters, aerogels, amorphous solids, and unusual phase
transition in the Ising model. Among various regular fractals, the
Vicsek fractals~\cite{Vi83} are a class of typical candidates for
exact mathematical ones and have received much attention. A variety
of structural and dynamical properties of Vicsek fractals have been
investigated in much detail, including eigenvalue
spectrum~\cite{JaWu94}, eigenstates~\cite{WeGr85}, Laplacian
spectrum~\cite{BlJuKoFe03}, random walks~\cite{Vi84},
diffusion~\cite{WaLi92}, and so on. The results of these
investigations uncovered many unusual and exotic features of Vicsek
fractals.

A central issue in the study of complex systems is to understand how
their dynamical behaviors are influenced by underlying geometrical
and topological properties~\cite{Ne03,BoLaMoChHw06}. Among many
fundamental structural characteristics~\cite{CoRoTrVi07}, mean
geodesic distance is an important topological feature of complex
systems that are often described by graphs (or networks) where nodes
(vertices) represent the component units of systems and links
(edges) stand for the interactions between
them~\cite{AlBa02,DoMe02}. Mean geodesic distance is defined as the
mean length of the shortest paths between all pairs of nodes. It has
been well established that mean geodesic distance directly relates
to many aspects of real systems, such as signal integrity in
communication networks, the propagation of beliefs in social
networks or of technology in industrial networks. Recent studies
indicated that a number of other dynamical processes are also
relevant to mean geodesic distance, including disease
spreading~\cite{WaSt98}, random walks~\cite{CoBeTeVoKl07},
navigation~\cite{Robe0607}, to name but a few. Thus far great
efforts have been made to valuate and understand the mean geodesic
distance of different
systems~\cite{ChLu02,CoHa03,HoSiFrFrSu05,DoMeOl06,ZhChZhFaGuZo08,ZhCoFeRaRoZh08}.

Despite the importance of this structural property, to the best of
our knowledge, the rigorous computation for the mean geodesic
distance of Vicsek fractals has not been addressed. To fill this
gap, in this present paper we investigate this interesting quantity
analytically. We derive an exact formula for the mean geodesic
distance characterizing the Vicsek fractals. The analytic method is
on the basis of an algebraic iterative procedure obtained from the
self-similar structure of Vicsek fractals. The obtained precise
result shows that the mean geodesic distance exponentially with the
number of nodes. Our research opens the way to theoretically study
the mean geodesic distance of regular fractals and deterministic
networks~\cite{BaRaVi01,DoGoMe02,ZhZhZoChGu07}. In particularly, our
exact solution gives insight different from that afforded by the
approximate solution of stochastic fractals.

\begin{figure}
\begin{center}
\includegraphics[width=.9\linewidth,trim=100 0 100 0]{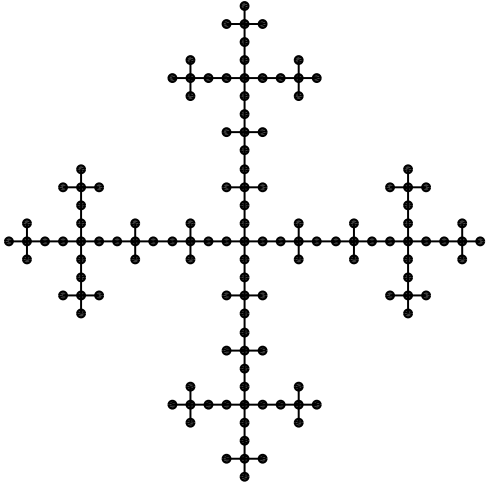}
\caption{Illustration of a particular Vicsek fractal $V_{3,2}$.}
\label{net}
\end{center}
\end{figure}

The classical Vicsek fractals are constructed
iteratively~\cite{Vi83,BlJuKoFe03}. We denote by $V_{f,t}$ ($t\geq
0$, $f\geq 2$) the Vicsek fractals after $t$ generations. The
construction starts from ($t=0$) a star-like cluster consist of
$f+1$ nodes arranged in a cross-wise pattern, where $f$ peripheral
nodes are connected to a central node. This corresponds to
$V_{f,0}$. For $t\geq 1$, $V_{f,t}$ is obtained from $V_{f, t-1}$.
To obtain $V_{f,1}$, we generate $f$ replicas of $V_{f,0}$ and
arrange them around the periphery of the original $V_{f,0}$, then we
connect the central structure by $f$ additional links to the corner
copy structures. These replication and connection steps are repeated
$t$ times, with the needed Vicsek fractals obtained in the limit $t
\rightarrow \infty$, whose fractal dimension is $\frac{\ln
(f+1)}{\ln3}$. In Fig.~\ref{net}, we show schematically the
structure of $V_{3,2}$. According to the construction algorithm, at
each time step the number of nodes in the systems increase by a
factor of $f+1$, thus, we can easily know that the total number of
nodes (network order) of $V_{f,t}$ is $N_{t}= (f+1)^{t+1}$.

After introducing the Vicsek fractals, we now investigate
analytically the mean geodesic distance between all the node pairs
in the fractals. We represent all the shortest path lengths of
$V_{f,t}$ as a matrix in which the entry $d_{ij}$ is the geodesic
distance from node $i$ to node $j$, where geodesic distance is the
path connecting two nodes with minimum length. The maximum value
$D_{t}$ of $d_{ij}$ is called the diameter of $V_{f,t}$. A measure
of the typical separation between two nodes in $V_{f,t}$ is given by
the mean geodesic distance $L_{t}$ defined as the mean of geodesic
lengths over all couples of nodes:
\begin{equation}\label{apl01}
  L_{t} = \frac{S_t}{N_t(N_t-1)/2}\,,
\end{equation}
where
\begin{equation}\label{total01}
  S_t = \sum_{i \in V_{f,t},\, j \in V_{f,t},\, i \neq j} d_{ij}
\end{equation}
denotes the sum of the geodesic distances between two nodes over all
pairs.
\begin{figure}
\begin{center}
\includegraphics[width=.9\linewidth,trim=100 0 100 0]{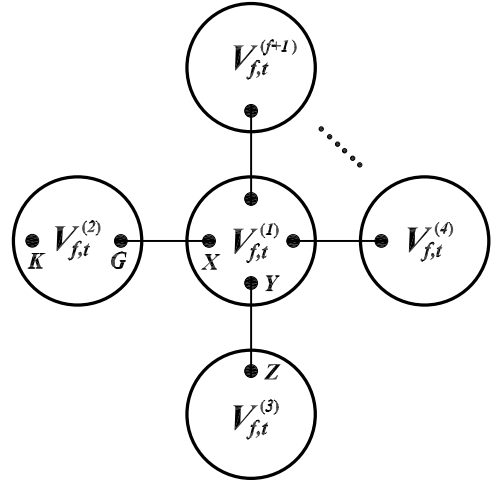}
\caption{A schematic illustration of the iterative construction for
$V_{f,t+1}$, which is obtained by joining $f+1$ copies of $V_{f,t}$
denoted as $V_{f,t}^{(1)}$, $V_{f,t}^{(2)}$, $\cdots$,
$V_{f,t}^{(f)}$, and $V_{f,t}^{(f+1)}$, respectively.} \label{copy}
\end{center}
\end{figure}

We continue by exhibiting the procedure of the determination of the
total distance and present the recurrence formula, which allows us
to obtain $S_{t+1}$ of the $t+1$ generation from $S_{t}$ of the $t$
generation. By construction, the fractal $V_{f,t+1}$ is obtained by
the juxtaposition of $f+1$ copies of $V_{f,t}$ that are
consecutively labeled as $V_{f,t}^{(1)}$, $V_{f,t}^{(2)}$, $\cdots$,
$V_{f,t}^{(f+1)}$, see Fig.~\ref{copy}. This obvious self-similar
structure allows us to calculate $S_t$ analytically. It is easy to
see that the total distance $S_{t+1}$ satisfies the recursion
relation
\begin{equation}\label{total02}
  S_{t+1} = (f+1)\, S_t + \Theta_t,
\end{equation}
where $\Theta_t$ is the sum over all shortest path length whose
endpoints are not in the same $V_{f,t}$ branch. The solution of
Eq.~(\ref{total02}) is
\begin{equation}\label{total03}
  S_t = (f+1)^{t}\, S_0 + \sum_{m=0}^{t-1} \left[(f+1)^{t-m-1} \Theta_m\right].
\end{equation}
Thus, all that is left to obtain $S_t$ is to compute $\Theta_m$.

The paths that contribute to $\Theta_t$ must all go through at least
one of the $2f$ edge nodes (such as $G$, $X$, $Y$, and $Z$ in
Fig.~\ref{copy}) at which the different $V_{f,t}$ branches are
connected. The analytical expression for $\Theta_t$, named the
crossing path length, can be derived as below.

Denote $\Theta_t^{\alpha,\beta}$ as the sum of all shortest paths
with endpoints in $V_{f,t}^{(\alpha)}$ and $V_{f,t}^{(\beta)}$. For
convenience, we denote by $V_{f,t}^{(1)}$ the central branch of
$V_{f,t+1}$. According to whether or not the two branches are
adjacent, we sort the crossing path length $\Theta_t^{\alpha,\beta}$
into two classes: $\Theta_t^{1,\phi}$ ($\phi >1$),
$\Theta_t^{\varphi,\theta}$ ($\varphi >1$, $\theta
>1$, and $\varphi \neq \theta$). For any two
crossing paths in the same class, they have identical length.
Therefore, in the following computation of $\Theta_t$, we will only
consider $\Theta_t^{1,2}$ and $\Theta_t^{2,3}$. The total sum
$\Theta_t$ is then given by
\begin{equation}\label{cross01}
\Theta_t =f\times\Theta_t^{1,2} + \binom {f}{2}\times\Theta_t^{2,3}.
\end{equation}

To calculate the crossing path length $\Theta_t^{1,2}$ and
$\Theta_t^{2,3}$, we give the following definition and notations. We
define external nodes of $V_{f,t}$ as the nodes that will be linked
to one of its copes at step $t+1$ to form $V_{f+1,t}$. Let $d_{t}$
denote the sum of length of the path from an external node of
$V_{f,t}$ to all nodes in $V_{f,t}$ including the external node
itself. We assume that the two branches $V_{f,t}^{(1)}$ and
$V_{f,t}^{(2)}$ are connected at two nodes $X$ and $G$, which
separately belong to $V_{f,t}^{(1)}$ and $V_{f,t}^{(2)}$, and that
$V_{f,t}^{(1)}$ and $V_{f,t}^{(3)}$ are linked to each other at two
nodes $Y$ and $Z$ that are in $V_{f,t}^{(1)}$ and $V_{f,t}^{(3)}$,
respectively.

In order to determine $d_{t}$, we should compute the diameter
$D_{t}$ of $V_{f,t}$ first. By construction, one can see that the
diameter $D_{t}$ equals the path length between arbitrary pair of
external nodes of $V_{f,t}$. Thus, we have the following recursive
relation:
\begin{equation}\label{diam01}
D_{t+1} =3\,D_{t}+2.
\end{equation}
Considering the initial condition $D_{0}=2$, Eq.~(\ref{diam01}) is
solved inductively to obtain
\begin{equation}\label{diam02}
D_{t}=3^{t+1}-1,
\end{equation}
which is independent of $f$.

We now calculate the quantity $d_{t+1}$. Let $K$ denote the external
node of $V_{f,t+1}$, which is in the branch $V_{f,t}^{(2)}$. By
definition, $d_{t+1}$ can be given by the sum
\begin{eqnarray}\label{dist01}
d_{t+1}&=&\sum_{j\in V_{f,t+1}} d_{Kj}\nonumber \\
&=&\sum_{u \in V_{f,t}^{(2)}} d_{Ku}+\sum_{v \in
V_{f,t}^{(1)}}d_{Ku}+(f-1)\sum_{w \in V_{f,t}^{(3)}}d_{Kw}\nonumber \\
&=&d_t+\sum_{v \in V_{f,t}^{(1)}}d_{Ku}+(f-1)\sum_{w \in
V_{f,t}^{(3)}}d_{Kw}.
\end{eqnarray}
We denote the second and third terms in Eq.~(\ref{dist01}) by $g_t$
and $q_t$, respectively. Thus, $d_{t+1}=d_t+g_t+q_t$. The quantity
$g_t$ is evaluated as follows:
\begin{eqnarray}\label{dist02}
g_{t}&=&\sum_{v \in V_{f,t}^{(1)}}(d_{KG}+d_{GX}+d_{Xu})\nonumber \\
&=&d_t+N_t\times(D_{t}+1),
\end{eqnarray}
where $d_{KX}=D_{t}$ and $d_{GX}=1$ were used. Analogously,
\begin{eqnarray}\label{dist03}
q_{t}&=&(f-1)\sum_{w \in V_{f,t}^{(3)}}(d_{KG}+d_{GX}+d_{XY}+d_{YZ}+d_{Zw})\nonumber \\
&=&(f-1)\left[d_t+N_t\times2(D_{t}+1)\right].
\end{eqnarray}
With Eqs.~(\ref{dist02}) and ~(\ref{dist03}), Eq.~(\ref{dist01})
becomes
\begin{equation}\label{dist04}
d_{t+1}=(f+1)d_t+(2f-1)\times N_t\times(D_{t}+1).
\end{equation}
Using $N_t=(f+1)^{t+1}$, $D_{t} =3^{t+1}-1$ and $d_0=2f-1$,
Eq.~(\ref{dist04}) is resolved by induction
\begin{equation}\label{dist05}
d_{t}=\frac{1}{2} (2 f-1) \left(3^{t+1}-1\right) (1+f)^t.
\end{equation}

With above obtained results, we can determine the length of crossing
paths $\Theta_t^{1,2}$ and $\Theta_t^{2,3}$, which can be expressed
in terms of the previously explicitly determined quantities. By
definition, $\Theta_t^{1,2}$ is given by the sum
\begin{eqnarray}\label{cross02}
\Theta_t^{1,2} &=& \sum_{i \in V_{f,t}^{(1)},\,\,j\in V_{f,t}^{(2)}} d_{ij}\nonumber \\
&=& \sum_{i \in V_{f,t}^{(1)},\,\,j\in V_{f,t}^{(2)}}
(d_{iX}+d_{XG}+d_{Gj})\nonumber \\
&=& N_t\,\sum_{i \in V_{f,t}^{(1)}} d_{iX}+(N_t)^2+N_t\,\sum_{j\in
V_{f,t}^{(2)}}d_{Gj}\nonumber \\
&=& 2N_t\,\sum_{i \in V_{f,t}^{(1)}} d_{iX}+(N_t)^2,
\end{eqnarray}
where we have used the equivalence relation $\sum_{i \in
V_{f,t}^{(1)}} d_{iX}=\sum_{j\in V_{f,t}^{(2)}}d_{Gj}$.

Proceeding similarly,
\begin{eqnarray}\label{cross03}
\Theta_t^{2,3} &=& \sum_{i \in V_{f,t}^{(2)},\,\,j\in V_{f,t}^{(3)}} d_{ij}\nonumber \\
&=& \sum_{i \in V_{f,t}^{(2)},\,\,j\in V_{f,t}^{(3)}}
(d_{iG}+d_{GX}+d_{XY}+d_{YZ}+d_{Zj})\nonumber \\
&=&2 N_t\,\sum_{i \in V_{f,t}^{(2)}}
d_{iG}+2\,(N_t)^2+(N_t)^2\,d_{XY}\nonumber \\
&=& 2N_t\,\sum_{i \in V_{f,t}^{(2)}} d_{iG}+(N_t)^2(D_{t}+2).
\end{eqnarray}
Inserting Eqs.~(\ref{cross02}) and~(\ref{cross03})  into
Eq.~(\ref{cross01}), we have
\begin{equation}\label{cross04}
\Theta_t=(f^2+f)N_t\,d_t+f^2\,(N_t)^2+\frac{f(f-1)}{2}(N_t)^2\,D_{t}.
\end{equation}
Substituting Eq.~(\ref{cross04}) into Eq.~(\ref{total03}) 
and using the initial value $S_0=f^2$, 
we can obtain the exact expression for the total distance
\begin{eqnarray}
S_t&=&\frac{1}{6 f+4}(f+1)^t \Big[
 f^2 (f+1)^t\left(3^{t+1}+1\right)\nonumber \\&+&3
\left(3^{t+1}-1\right) f^3 (f+1)^t+4 \left((f+1)^t-1\right)\nonumber
\\&-&2 f \left(3^{1+t} (f+1)^t-4 (1+f)^t+1\right)\Big]
\end{eqnarray}
Then the analytic expression for mean geodesic distance can be
obtained as
\begin{eqnarray}\label{apl02}
L_{t} &=&\frac{1}{(3 f^2+5f+2)[(f+1)^t-1]} \Big[ f^2
(f+1)^t\left(3^{t+1}+1\right)\nonumber \\&+&3 \left(3^{t+1}-1\right)
f^3 (f+1)^t+4 \left((f+1)^t-1\right)\nonumber
\\&-&2 f \left(3^{1+t} (f+1)^t-4 (1+f)^t+1\right)\Big].
\end{eqnarray}
In the infinite system size, i.e., $t\rightarrow \infty$
\begin{eqnarray}\label{apl03}
L_{t}\sim 3^{t+1}= (N_t)^\frac{\ln3}{\ln (f+1)},
\end{eqnarray}
where the exponent $\frac{\ln3}{\ln (f+1)}$ is equal to the
reciprocal of the fractal dimension. Thus, the mean geodesic
distance grows exponentially with increasing size of the system. In
contrast to many recently studied network models mimicking real-life
systems in nature and society~\cite{AlBa02,DoMe02}, the Vicsek
fractals are not small worlds despite of the fact that these
fractals show similarity (fractality) observed in many real-world
systems.

\begin{figure}
\begin{center}
\includegraphics[width=\columnwidth]{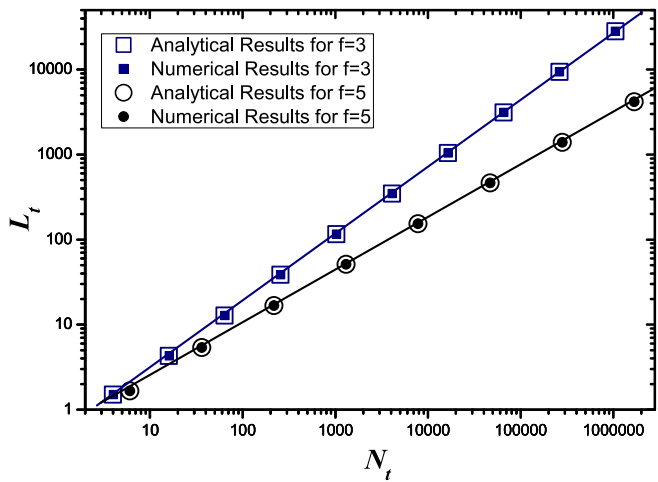}
\end{center}
\caption[kurzform]{\label{distance} Mean geodesic distance $L_{t}$
versus network order $N_{t}$ on a log-log scale. The solid lines are
guides to the eyes.}
\end{figure}

We have checked our analytic result against numerical calculations
for different $f$ and various $t$. In all the cases we obtain a
complete agreement between our theoretical formula and the results
of numerical investigation, see Fig.~\ref{distance}.

To sum up, in complex systems the mean geodesic distance plays an
important role. It has a profound impact on a variety of crucial
fields, such as information processing, disease or rumor
transmission, network designing and optimization. In this paper, we
have derived analytically the solution for the mean geodesic
distance of Vicsek fractals which have been attracting much research
interest. We found that in the infinite network size limit the mean
geodesic distance scales exponentially with the number of nodes. Our
analytical technique could guide and shed light on related studies
for deterministic fractals and network models by providing a
paradigm for calculating the mean geodesic distance. Moreover, as a
guide to and a test of approximate methods, we believe our vigorous
solution can prompt the studies on random fractals.

We would like to thank Yichao Zhang for preparing this manuscript.
This research was supported by the National Basic Research Program
of China under grant No. 2007CB310806, the National Natural Science
Foundation of China under Grant Nos. 60496327, 60573183, 90612007,
60773123, and 60704044, the Shanghai Natural Science Foundation
under Grant No. 06ZR14013, the China Postdoctoral Science Foundation
funded project under Grant No. 20060400162, the Program for New
Century Excellent Talents in University of China (NCET-06-0376), and
the Huawei Foundation of Science and Technology (YJCB2007031IN).

\end{document}